\documentclass[12pt,twoside]{article}
\usepackage{epsfig}
\usepackage{amssymb,amsbsy,amsmath}
\newcommand{\mycaption}[1]{%
\begin{center}\begin{minipage}{125mm}%
\caption[]{#1}%
\end{minipage}\end{center}%
}%
\begin{document}
%
\title{FERMIONIC MOLECULAR DYNAMICS:
MULTIFRAGMENTATION IN HEAVY--ION COLLISIONS AND IN EXCITED
NUCLEI}
 
\author{H. Feldmeier\thanks{email: h.feldmeier\char'100gsi.de,
              WWW:~http://www.gsi.de/$\sim$feldm}\;
and J. Schnack\thanks{email: j.schnack\char'100gsi.de,
              WWW:~http://www.gsi.de/$\sim$schnack}\\[2mm]
Gesellschaft f\"ur Schwerionenforschung mbH, \\ 
Postfach 110 552, D-64220 Darmstadt}

\date{}

\maketitle

\section{INTRODUCTION AND SUMMARY}
How multifragmentation happens in heavy ion collisions is still
a matter of debate. Explanations reach from nucleation over self
organization, spinodal decomposition to cold break--up and
survival of initial correlations. For an overview see
ref. \cite{Hir94}. A key question is the time scale of
the reaction.  Slow processes like nucleation or self
organization are hindered if the expansion of the whole system
is too fast. Another issue is the relaxation time for thermal
and chemical equilibrium which is important when statistical
models are used to explain multifragmentation
\cite{BBI95,Gro90,Fri88,PaN95}. 

If one considers the decay of excited spectator matter which has
not been compressed, the expansion preceding the fragmentation
might be slow enough to allow for global equilibration. However,
below excitations energies of about 6 MeV per nucleon an
equilibrated nucleus will most likely cool down by evaporating
nucleons but not by breaking into many intermediate mass
fragments (IMF, $Z \ge 3$). The reason is that the barrier for
multifragmentation is too high and the nucleus has cooled already
by evaporation before it fragments into pieces.  In order to
drive an expansion across the barrier into the spinodal region
just by means of thermal excitation at normal density, one needs
excitation energies between 6 and 10 MeV per particle
\cite{PaN95}.  
For these energies the question arises: Can in a heavy--ion
collision, which chops off a large fraction of the nucleus, the
excitation of the remaining spectator part be distributed fast
enough among the kinetic degrees of freedom such that the
expansion sets in only after thermal equilibration? Or is it
more likely that a peripheral or semi--peripheral collision
creates a non--equilibrium object with strong local fluctuations,
which drive the system across the barrier towards
multifragmentation right away without first thermalizing and
then expanding? In such a situation a common mean field would
not establish anymore and the system could not be regarded as an
equilibrated Fermi gas in a mean field.

We want to argue here that a non--thermal situation is actually
very helpful in getting the nucleus across the barrier for
multifragmentation which exists in the equilibrium
potential--energy surface.  
%
%
%
%
On the way to equilibrium the system can then easily cross the
equilibrium barriers or is already behind them.  A non--equilibrium system
can feed into all parts of the coexistence region in the phase
diagram, large and small volumes, i.e. many or few
fragments. And thus the isotopic ratios in the ensemble can
reflect the properties of the coexistence region and the
liquid--gas phase transition \cite{Poc95}.  Therefore, we see the
possibility that the phase diagram and the liquid--gas phase
transition can also be investigated by non--equilibrium
multifragmention of a wounded spectator and may be even better
than by thermally excited nuclei (hot Fermi gas
in a mean--field) which go through thermal expansion and
subsequent formation of fragments.

For the participant matter created for example in central
collisions the compression is much stronger and the excitation
energy much higher. A transient pressure exists during the time
of instreaming matter and thereafter due to recoil of promptly
emitted particles. This together with the short mean free path
at high excitations is in favour of multifragmentation
originating from a more thermalized source.  But the whole
system is expanding and cooling fast so that it is questionable
if there are enough collisions to ensure local equilibrium until
freeze out. Experiments show for example that a large part of
the excitation energy is converted into radial flow
\cite{Rei96}.
 
There is little hope to decide upon these questions
experimentally in a unique way, because it 
is very difficult to measure temperature and flow profiles 
\cite{Poc95,Rei96,NHW95} and even harder or impossible to infer
experimentally on the time scale of the evolution of the
system. Therefore, microscopic transport models which do not
assume equilibration are needed for a better understanding. These
models should go beyond the mean field approach, which is a kind
of equilibrium assumption in itself, so that in principle they
are capable to describe many--body correlations like the
formation of fragments.  QMD, AMD and FMD are molecular dynamics
models which assert this claim. How equilibrium is achieved can
then be studied by comparing distributions, for example of mass,
charge, kinetic energy etc, with equilibrium distributions.

In the following two sections we investigate within Fermionic
Molecular Dynamics (FMD) the decay of a compound system with 46
or 80 nucleons which was created in a heavy--ion collision at a
beam energy of about 35~$A$MeV and the decay of $^{56}$Fe which
we put in an excited state by scaling the whole many--body wave
function and/or randomly moving the centroids of the wave
packets. For the definition of the model see ref.
\cite{Fel90,FBS95,FeS95,FeS97,ScF97}.

The succeeding section shows FMD collisions of $^{19}$F +
$^{27}$Al and $^{40}$Ca + $^{40}$Ca in the Fermi energy domain
where multifragmentation is the dominant reaction mechanism. The
system, however, does not go through a thermalized situation.  In
section 'Decay of excited nuclei' FMD evolutions of randomly
excited nuclei (artificially thermalized source inside the
multifragmentation barrier) are investigated.  They do not show
multifragmentation within a set of about 20 runs. Either they
vapourize into individual nucleons, or after expanding and
blowing off outer layers the inner part contracts again and an
evaporation residue, which can be rather small, is left over.
Only if not all correlations are destroyed the excited nucleus
expands and decays into fragments and single nucleons, quite
similar to the decay following the collision.

\section{MULTIFRAGMENTATION IN COLLISIONS}

Central and semi--peripheral collisions of $^{19}$F + $^{27}$Al 
and  $^{40}$Ca + $^{40}$Ca calculated within FMD are shown in
fig. 1 and 2. We choose an energy of
32 $A$MeV and 35 $A$MeV for which the relative velocity between the two
nuclei is about the Fermi velocity. Here we expect the break
down of the mean--field picture which prevails in the dissipative
regime (up to about $E_{beam}=15 A$MeV)
where the system either fuses or undergoes a strongly damped
binary collision \cite{FeS97}. When the collective velocity becomes
comparable to the internal velocities of the nucleons, a common
mean field cannot be established any longer and non--equilibrium
effects will be important. The picture of a hot Fermi gas in a mean
field will no longer be true.

The following figures show a variety of events as contour plots
of the one--body density in coordinate space. This density is
integrated over the $z$--direction. Figure 1 and
2 present runs at two impact parameters and two initial
orientations of the intrinsically deformed ground states.

In fig. 1 one sees for both impact parameters the creation
of a source which lives for about 100 fm/c before it fragments
into pieces of all sizes. The time is given in the upper right
corners. For the larger impact parameter the source is more
stretched and the final momenta of the fragments are not as
isotropic as in central collision.

For the larger system $^{40}$Ca + $^{40}$Ca (fig. 2) the
situation is similar except that the number of IMFs is larger
and at impact parameter $b=2.75$ fm the outgoing fragments have
still a more isotropic momentum distribution. Pronounced flow
sets in for larger impact parameters.

Although the one--body densities at $t=100$ fm/c in fig. 1 
and at $t=120$ fm/c in fig. 2 look very thermalized, they
are not. There are still strong many--body correlations which
just cannot be seen in a one--body distribution. Analyzing the
time evolution of a cluster one sees that the correlations
between the wave packets which finally compose the fragment can
be followed back for rather long time. The heavy--ion collision
does not completely randomize the many--body state. The
important role of correlations for multifragmentation in FMD
will be discussed further in the following section where we
destroy these correlations artificially.

\begin{figure}[tttt]
\unitlength1mm
\begin{picture}(140,155)
%
%
\put(0,10){\epsfig{file=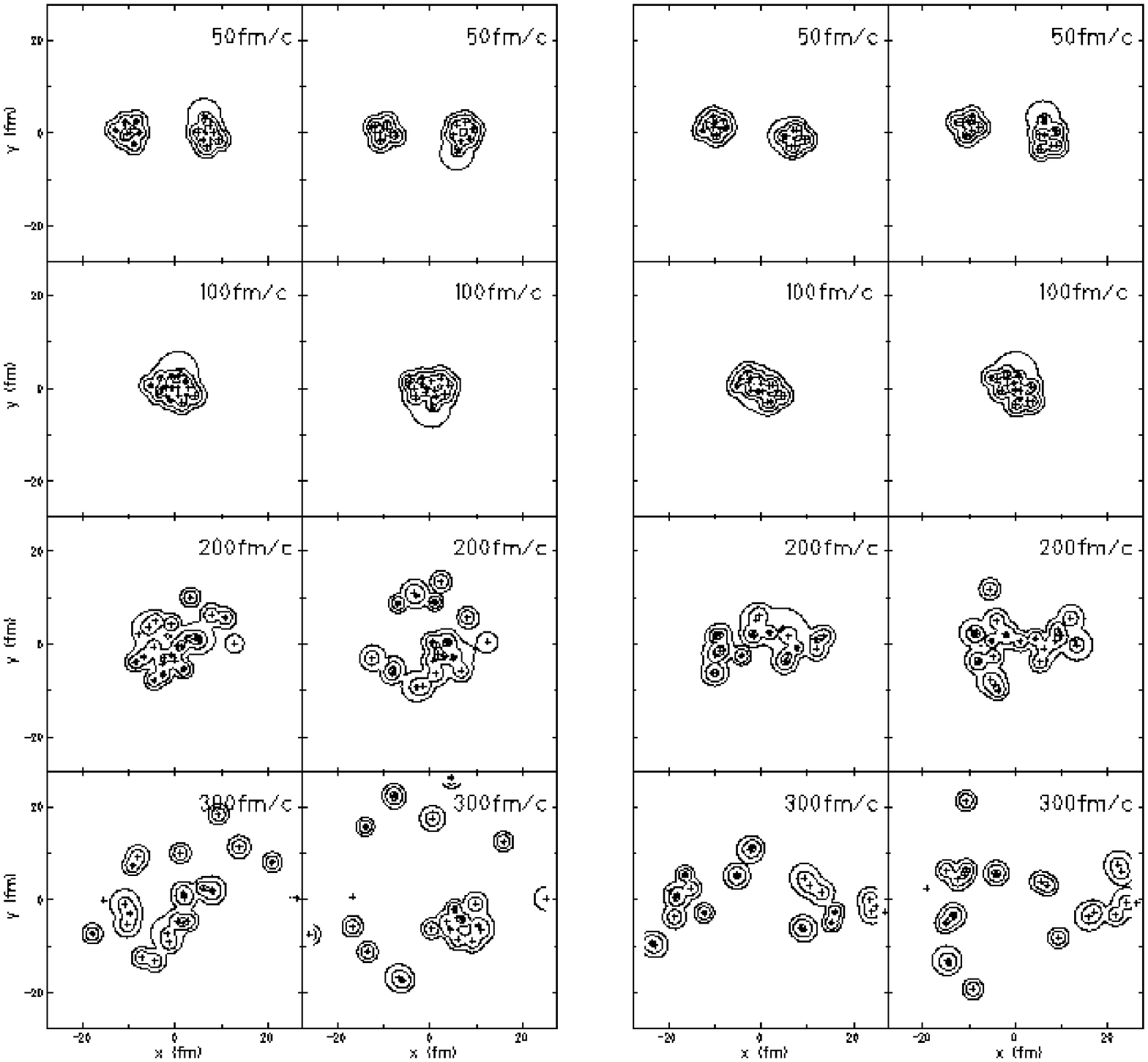,width=150mm}}
\end{picture}
\mycaption{One--body density in coordinate space integrated
over~$z$ for 
$^{19}$F+$^{27}$Al collisions at $32~A$MeV,
$b=0.5~$fm (l.h.s.) and $b=2.5~$fm (r.h.s.).
The contour lines depict the density at
0.01,~0.1,~0.5~fm$^{-2}$. Crosses indicate the mean positions
$\vec{r}_k$ of the wave packets.
}
\end{figure} 

\begin{figure}[tttt]
\unitlength1mm
\begin{picture}(140,155)
%
%
\put(0,10){\epsfig{file=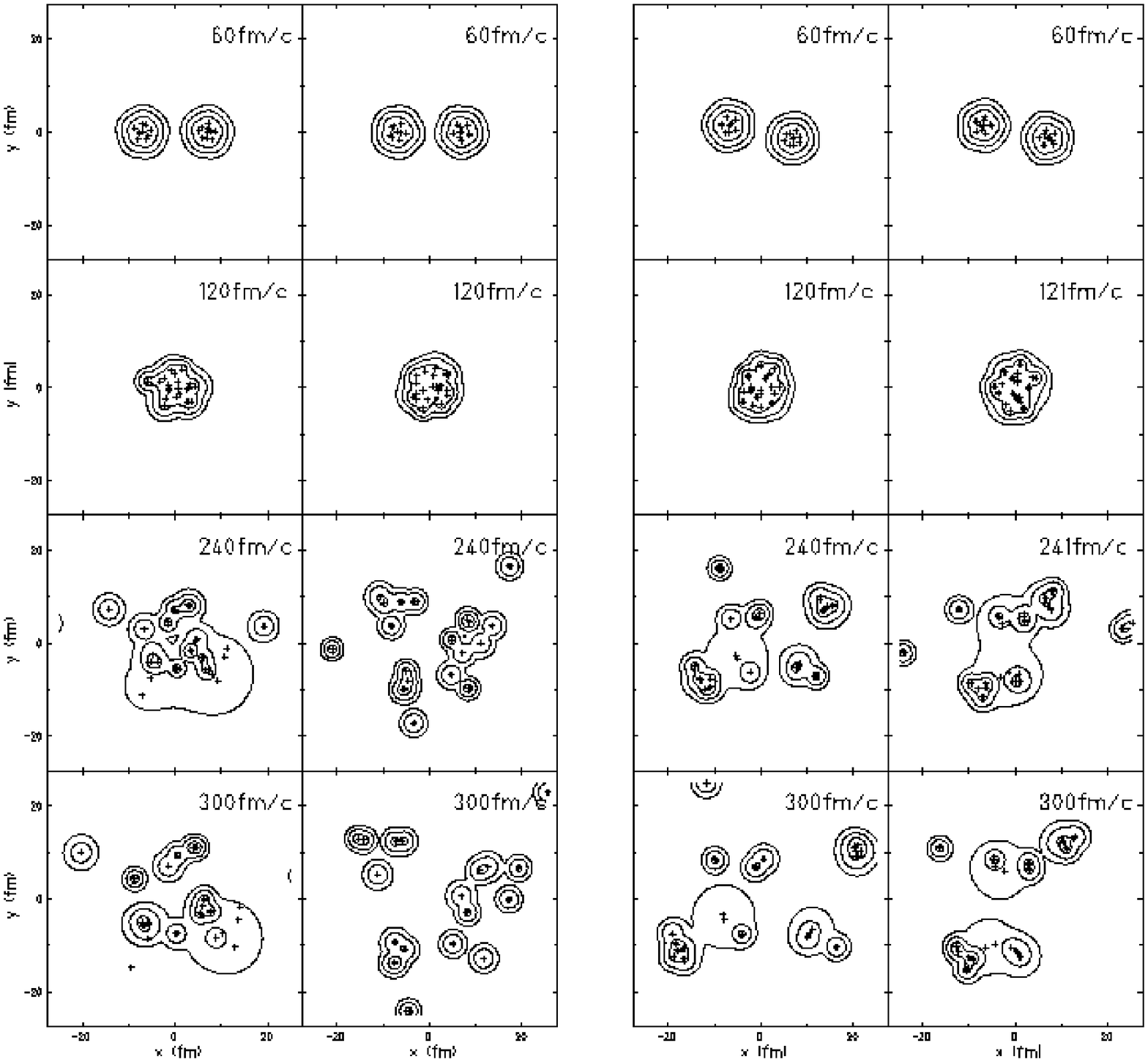,width=150mm}}
\end{picture}
\mycaption{Same as fig. 1 but for 
$^{40}$Ca+$^{40}$Ca collisions at $35~A$MeV,
$b=0.25~$fm (l.h.s) and $b=2.75~$fm (r.h.s).
}
\end{figure} 

\section{DECAY OF EXCITED NUCLEI}

In order to study the influence of many--body correlations on
the decay pattern of an excited nuclear system we create in this
section various initial states by exciting a $^{56}$Fe nucleus
through a combination of scaling the ground--state density and
randomly displacing the mean positions of the wave packets
without changing the density.  Both create excitation, but while
the scaling does not destroy the inter--particle correlations,
the random displacement does.

In the FMD ground state the wave packets arrange in phase space
such that the total energy is minimized. This ground state is an
intrinsic state in which the relative positions and orientations
of the wave packets in coordinate and momentum space reflect
many--body correlations. If one destroys these correlations by
randomly displacing all mean position parameters $\vec{r}_k$
around their original positions within a small circle of radius
0.2~fm perpendicular to $\vec{r}_k$ the $^{56}$Fe nucleus
achieves already 1.6~MeV excitation energy per particle,
although the one--body density remains practically unchanged. The
first row in fig. 3 shows the time evolution of an excited state
with 9.7~$A$MeV excitation energy which was achieved by random
displacements within 2~fm. One sees a fast expansion of the
outer density layers which carries away a lot of energy and the
survival of a residue in the center which evaporates nucleons on
a longer time scale (not shown here).

In the second row the same magnitude of random displacements leads to
11.2~$A$MeV excitation energy. Here the expansion is faster and
the residue is smaller. But there are also other cases around 11.2~$A$MeV
where no residue is left over. Already above $E^*=12~A$MeV no residues
are observed anymore and the nucleus vaporizes
completely. There is a sharp transition around $E^*=11~A$MeV
where the thermal expansion can not be brought to a halt by the
attractive interaction anymore.

In the fourth row we scaled the density in coordinate space by a
factor 2.2, which implies a scaling of the momentum density by
1/2.2, and in addition displaced the mean position vectors
within 0.5~fm. The result is an excitation energy of
$E^*=11.3~A$MeV similar to the randomized case in the second
row. Up to 200~fm/c also the density develops rather similar to
this case. The main difference to random excitation is that the
compressed nuclei (by scaling) vaporize mainly by spreading of
the wave packets and less by radial motion of the centroids
(crosses in fig. 3) whereas in the randomized nuclei the
centroids move out faster. At lower excitation energies,
i.e. below 10~$A$MeV (not shown here), a few particles are
emitted fast, then the residue goes through damped monopole
vibrations while thermalizing, and finally evaporates nucleons
after some delay.  From this and other runs we conclude that
concerning multifragmentation the coherent compression does not
change the picture too much compared to random excitation.

The last row shows the evolution of a system in which only the mean
position vectors are scaled down by 0.6, but the width parameters
and the mean momentum parameters are not changed. On top of that
a random displacement within 0.3~fm was applied. The resulting excitation
energy $E^*=10~A$MeV is comparable with the cases discussed
above. But unlike in the first row the system expands without
forming a residue in the centre and then undergoes
multifragmentation. The reason is that 
this way of exciting the nucleus does not destroy as much the spatial
correlations. Those wave packets which are grouped together
in the ground state will still be close after the
excitation and the correlations survive to a large extend the
expansion. The density develops wrinkles rather early (see frame
at 50~fm/c) which rapidly become cracks (see 100~fm/c) and at
200~fm/c, where in the first row the randomized system is still
rather compact, the fragments are widely separated. 
One should note that although the one--body density of the
initial state seems perfectly symmetric and no extra momentum
has been given to the wave packets the expansion amplifies the
correlations and the initial symmetry of the one--body density
is broken by rapidly growing fluctuations.

This shows that in FMD many--body correlations play an important
role in the formation of fragments. It seems that the time,
which is set by the expansion, is too short to allow the
many--body state to develop the special kind of many--body
correlations which are needed to make a cluster with low enough
excitation energy, so that it can survive.

The absence of multifragmentation in other calculations of the
FMD type \cite{KiD96,CCG96} is interesting. Especially in
\cite{KiD96} the effect of decoupling the center--of--mass
degrees of freedom from the internal degrees of freedom of a
fragment was investigated because in a many--body state, which is a
(antisymmetrized) product of single--particle packets, a fragment
has always about 10~MeV kinetic energy in the localization of
the c.m.--coordinate, which for small fragments might decrease
the production probability \cite{OHM92}. Even with the
additional degrees of freedom no multifragmentation was found in
ref. \cite{KiD96}. 

One must however be very careful in comparing
different FMD type calculations because we found that the result
of a heavy--ion reaction is very sensitive to the two--body Hamiltonian
employed. In particular if the ground state properties of the
different fragments are not described well (with the very same
Hamiltonian) the outcome of a collision is unpredictable. But
even for interactions which reproduce ground states equally well
it can happen that for example two light nuclei do not fuse
anymore at low energy (a must for any model). Therefore one may
not yet conclude that the FMD trial state is too much restricted
in its degrees of freedom since we see multifragmentation,
dissipative binary reactions and fusion with the same
interaction \cite{FeS97}. 

\begin{figure}[tttt]
\unitlength1mm
\begin{picture}(140,190)
%
%
\put(0,05){\epsfig{file=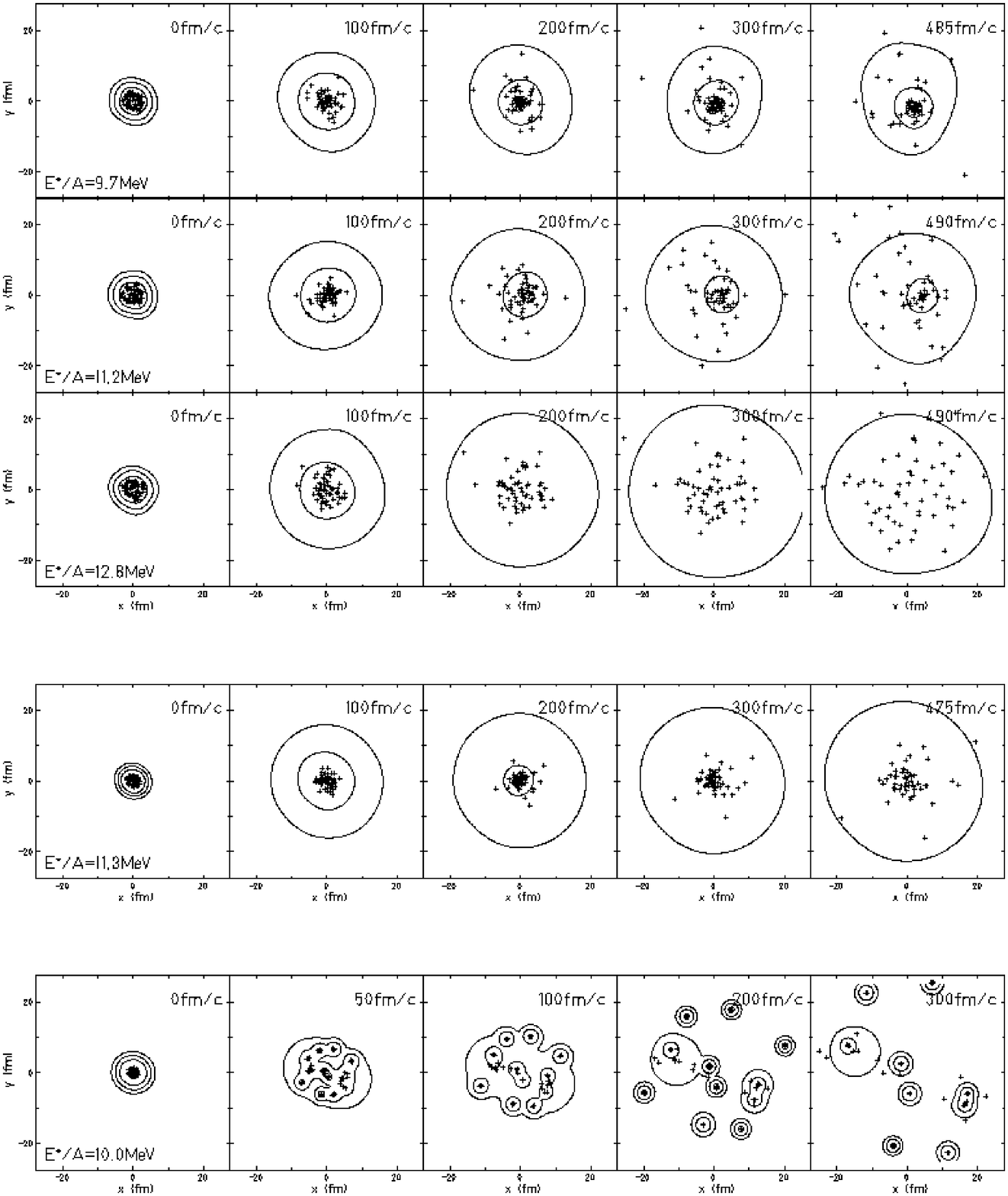,width=150mm}}
\end{picture}
\mycaption{Same as fig. 1 but for
$^{56}$Fe deexcitations at various excitations.
First three rows: random displacement of mean positions
$\vec{r}_k$ (crosses).
Fourth row: density scaled by 2.2 and small random
displacements.
Fifth row: $\vec{r}_k \rightarrow 0.6\vec{r}_k$ and small random
displacements.   
}
\end{figure} 

\end{document}